# BSE: A MINIMAL SIMULATION OF A LIMIT-ORDER-BOOK STOCK EXCHANGE

Dave Cliff

Department of Computer Science
University of Bristol
Bristol BS8 1UB, U.K.

csdtc@bristol.ac.uk

**ABSTRACT**
This paper describes the design, implementation, and successful use of the *Bristol Stock Exchange* (BSE) a novel minimal simulation of a centralized financial market, based on a Limit Order Book (LOB) such as is commonly in major stock exchanges. Construction of BSE was motivated by the fact that most of the world's major financial markets have automated, with trading activity that previously was the responsibility of human traders now being performed by high-speed autonomous automated trading systems. Research aimed at understanding the dynamics of this new style of financial market is hampered by the fact that no operational real-world financial exchange is ever likely to allow experimental probing of that market while it is open and running live, forcing researchers to work primarily from time-series of past trading data. Similarly, university-level education of the engineers who can create next-generation automated-trading systems requires that they have hands-on learning experiences in a sufficiently realistic teaching environment. BSE as described here addresses both needs: it has been successfully used for teaching and research in a leading UK university since 2012, and the BSE program code is freely available as open-source on *GitHub*.

Keywords: simulation for education, financial markets, automated trading.

## 1. INTRODUCTION

This paper reports on the design, implementation, and successful use of a simulated financial exchange, for use in university teaching and research. The simulated exchange implements the same central dynamic data structure (the *Limit Order Book,* or LOB) as is found in major national financial exchanges such as NYSE or NASDAQ in the USA, and LSE in the UK. Having been developed at the University of Bristol, the exchange simulator is named the *Bristol Stock Exchange* (BSE). BSE was deliberately written to be easy to use by, and to understand for, novice programmers; and it does not require powerful hardware to run (it happily executes, slowly, on the popular *Raspberry Pi* low-cost single-board computer). BSE is written in *Python* (v.2.7) and the full source-code has been made freely available as open-source on the *GitHub* repository.

The crucial difference between BSE and traditional financial-market simulators that operate by regurgitating a time-series database of historical transaction prices is that the traders in BSE can directly affect the prices at which transactions take place, pushing prices up when demand exceeds supply and driving prices down when supply exceeds demand. In BSE, the price at time $t + 1$ is not simply whatever price is recorded on the historical time series but is instead directly dependent on the actions and interactions of the traders active in the market at time $t$. This is explained and explored in more detail later in this paper.

The creation of BSE was motivated primarily by the observation that many of the world's major financial markets now have very high levels of automation, with human traders having been replaced by autonomous *algorithmic trading* systems, known colloquially as "robot traders" or simply as "algos". Early automated trading systems were often introduced to perform simple routine trading activities that it was not worth asking a highly-paid human trader to do, but over time the capabilities of robot traders, particularly the amount of data that they could assimilate and act upon, and/or the sheer speed at which they could respond to changes in the market, meant that the robot traders could outperform human traders, at lower cost. At that point, simple considerations of economic efficiency meant that robots traders proliferated in many markets, and the number of trades that involved human counterparties negotiating at the point of execution fell sharply. The change to automated trading has altered the dynamics of major financial markets and has created a demand for people with university-level education in the design and construction of automated trading systems, and in the analysis and management of automated markets. BSE is a simple minimal abstraction of an automated market, and includes reference implementations of a number of well-known robot trader algorithms. Section 2 of this paper gives background information on auctions, financial markets, and the LOB. Section 3 then describes BSE's LOB, its array of robot traders, and how the robot traders can be used to populate a market and interact via the LOB. After that, Section 4 talks about experiences of using BSE and its array of robot traders in teaching and research at the University of Bristol.

## 2. BACKGROUND

In very many human societies, for hundreds or thousands of years, buyers and sellers have met at marketplaces and haggled. When haggling, the seller states the offer-price that he or she wants to sell at, the buyer typically responds with a bid-price that is lower than the offer. The seller might drop the offer a little; the buyer might increase the bid a little; and they repeat these price revisions until they have struck a deal or one side walks away.

In the language of economics, the word "auction" is used to refer to the means by which buyers and sellers come together to agree a transaction price, exchanging money for goods or services. Haggling is a form of auction. There are lots of other different types of auction, here we will briefly review only five, but our review ends with the most economically significant style of auction, as is used in most of the world's major financial markets.

One well-known type of auction is the *English Auction*, where the seller stays silent and the buyers announce increasing bid-prices until only one buyer remains, who then gets the deal. This is a popular way of selling fine art, and livestock too. A more technical name for the English auction is a *first-price ascending-bid auction* because the first (highest) price becomes the transaction price. In contrast, if you've ever bought anything on eBay, you'll know that a lot of the auctions there are ascending-bid, but run as *second-price:* that is, you win the deal by bidding highest, but the price you pay is the bid-price of the second-highest bidder. And, when you're in a shopping mall, you're in an auction too. It's what economists refer to as a *Posted Offer Auction*: the sellers name (or "post") their offer-prices, and the buyers simply take it or leave it at that price.

However, if you go to Amsterdam or Rotterdam and try to buy tulip or daffodil bulbs (a big business in the Netherlands) you'll see almost exactly the opposite process in action. In the *Dutch Flower Auction*, the buyers stay silent while the seller starts with an initial high offer-price and then gradually drops the offer price until a buyer jumps in to take the deal. This is a *descending-offer* auction.

In many of the world's major financial markets, the style of auction used is a very close relative of the basic haggling process. This is much like having an ascending-bid and a descending-offer auction going on in the one market simultaneously. It is known as the *Continuous Double Auction*, or *CDA*. In the CDA, a buyer can announce a bid at any time and a seller can announce an offer at any time. We'll refer to bids and offers here collectively as *quotes*. While this is happening, any seller can accept any buyer's bid at any time; and any buyer can accept any seller's offer at any time. When a quote is accepted by a counterparty, the deal goes through and the quote's price becomes the transaction price for that deal. The CDA is a continuous asynchronous process, and it needs no centralized auctioneer, but it does need some way of recording the bids and offers that have been made and not yet transacted: this is the *limit order book* (LOB) that we will look at in some detail in this paper. In market terminology, a "limit order" is a quote that will only be executed when a counterparty is found who is willing to transact at the quote's pre-specified *limit price*: this distinguishes it from other types of order that execute immediately, at whatever price the market will bear at the moment the order is submitted.

The CDA interests economists because, even with a very small number of traders, the transaction prices rapidly approach the theoretical market equilibrium price. The equilibrium price is the price that best matches the quantity demanded to the quantity supplied by the market, and in that sense it is the most efficient price for the market. The CDA is also of pragmatic interest because of the trillions of dollars that flow through national and international CDA-based markets in commodities, equities (stocks and shares), foreign exchange, fixed income (tradable debt contracts such as government bonds, known in the UK as *gilts* and in the USA as *treasury bills*), and derivatives contracts. Although there are still some exchanges where human traders physically meet in a central trading pit and shout out verbal bids and offers, in very many major markets the traders engage with one another remotely, via a screen-based electronic market, interacting by placing quotes for specific quantities at specific prices on the LOB.

The LOB displays data that summarizes all the outstanding bids and offers, i.e. the "live" quotes that have not yet cancelled by the traders that originated them. In market terminology, offers are also referred to as *asks*, and the LOB has is typically described as having two sides: the *bid side* and the *ask side*. The bid side shows the prices of outstanding bid orders, and the quantity available at each of those prices, in descending order, so that the best (highest) bid is at the top of the book. The ask side shows the prices of outstanding asks, and the quantity available at each of those prices, in ascending order, so that the best (lowest) ask is at the top of the book.

So, for example, if there are two traders each seeking to buy 30 shares in company XYZ for no more than $1.50 per share, and one trader hoping to buy 10 for a price of $1.52; and at the same time if there was one trader offering 20 shares at $1.55 and another trader offering 50 shares at $1.62, then the LOB for XYZ would look like the one illustrated in Figure 1: traders would speak of XYZ being priced at "152-55", and the bid-offer *spread* is $0.03.

The information shown on a LOB is referred to as "Level 2" or "market depth" data. In contrast, "Level 1" market data shows only the price and size (quantity) for the best bid and ask, along with the price and size of the last recorded transaction, of the instrument being traded. Some people like to try their hand at "day trading" on their home PCs and they often operate with even more restricted data, such as the time-series of whatever price the instrument was last traded at, or the mid-price, the point between the current best bid and the best ask (so in this example, the mid-price of XYZ would be $1.535). The richer the data, the more expensive it is to purchase from a commercial provider of financial data. Full Level 2 data is routinely used by professional traders in investment banks and hedge funds, but research and development engineers in those institutions are famously much better resourced than meagerly-funded university academics. Buying access to records of Level 2 data in the quantities needed for teaching or research is typically prohibitively expensive for routine academic use.

Figure 1: Illustrative Limit Order Book (LOB): how the LOB for a fictional stock with ticker-symbol XYZ might be displayed on a trader's screen. Left-hand (pale-text) columns show the bid-side of the book, quantity then price, ordered in descending order of price; right-hand (dark text) columns show the ask-side, price then quantity, in ascending order of price. See text for further discussion.

However, even if such historical data was available for free, it does not allow us to study what is known as *market impact*, where the actions of a particular trader or group of traders moves the price of a tradeable instrument. For example, if a trader sells a very large amount of IBM shares, the increased supply will depress the subsequent price of IBM stock (assuming that all other things, especially the level of demand for IBM stock, remain the same). In a teaching or research context, a database of historical price time-series cannot exhibit market impact: if a trader decides, on the basis of the price of IBM stock at time *t*, to sell 100 million IBM shares, the price at time *t+1*, i.e. the next price in temporal sequence available from the historical database, will be unaffected by that sudden massive increase in supply of IBM stock; that is, there will be no market impact effect. And yet we want to explore understand market impact from a research perspective, and we want our students to experience their systems dealing with market impact too.

For these reasons, and on the basis of earlier experiences with having supervised PhD students and postdoctoral researchers constructing more sophisticated and complex simulations of financial markets for use in research and in teaching (see e.g.: Stotter *et al.*, 2013, 2014; De Luca, 2016), I designed and implemented a simple minimal simulation of a LOB-based financial market. Because I work at the University of Bristol, I gave this simulator the name *Bristol Stock Exchange*, or BSE.

## 3. BSE: THE BRISTOL STOCK EXCHANGE

BSE has been successfully used as a research tool, most recently in experimental studies of applying deep learning neural networks (DLNN) machine learning methods to create robot traders that adapt to changing market circumstances and which learn from their experience of the market, discussed further in Section 4. In such contexts it is sensible to talk about the user of BSE as an *experimenter* or *researcher*. BSE was originally developed as, and was successfully used as, a resource in teaching on a masters-degree course at the University of Bristol: see Cliff (2018) for further discussion of use of BSE in teaching. In teaching contexts, the user of BSE is more appropriately referred to as an *educator* (someone setting up BSE to be used in teaching, to achieve learning outcomes in students) or as a *student* (i.e., someone using BSE to expand their learning and understanding of contemporary financial market systems). In the text that follows, *user* will be used as a generic term for experimenter, researcher, teacher, and student, all working with BSE.

BSE has the following features:

- While real financial exchanges will typically simultaneously maintain LOBs for tens, hundreds, or thousands of types of tradeable item (i.e., different stocks, or different commodities), BSE has just one LOB, for recording limit orders in a single anonymous type of tradeable item.
- BSE allows for the user to control the specification of any of a wide range of dynamics of supply and demand in the BSE market.
- BSE includes a number of pre-coded robot trading algorithms drawn from the literature on automated trading over the past 30 years. This allows the user to explore the dynamics of LOB-based CDA markets without having to write their own robot-trading algorithms.
- BSE is deliberately written as a simple, intelligible, single-threaded minimal simulation, in the widely used programming language *Python* 2.7, so that students with only elementary programming experience can readily experiment with altering aspects of the existing system, or extending it by adding their own robot trading algorithms. The BSE souce-code has been available as open-source from the *Github* code repository since October 2012 and has been downloaded many times, with a number of individuals contributing uploads of their edits and extensions of the system.

In Section 3.1 the BSE LOB is described in more detail; Section 3.2 then gives a brief overview of the array of robot traders currently available in BSE; Section 3.3 then talks about how a market session can be organized, with some number of robot traders interacting in BSE via the LOB; Section 3.4 gives details of how the market's supply and demand are specified, by defining the flow of customer orders into the market.

## 3.1. The BSE Limit Order Book (LOB)

BSE is a minimal simulation of a financial exchange running a limit order book (LOB) in a single tradable security. It abstracts away or simply ignores very many complexities that can be found in a real financial exchange. In particular, a trader can at any time issue a new order, which immediately replaces any previous order that the trader had on the LOB: that is, any one trader can have at most one order on a LOB at any one time. Furthermore, as currently configured, BSE assumes zero latency in communications between the traders and the exchange, and also conveniently assumes that after any one trader issues an order that alters the LOB, then any transaction triggered by the update is immediately resolved and the updated LOB is distributed to all traders before any other trader can issue another order: that is, all LOB updates are assumed to take place at zero latency. In the days when all traders were humans, the speed at which computerized updates to the LOB could be executed were so much faster than human reaction times that transmission latencies along conventional wired or wireless telecoms links could safely be assumed to be zero, as in BSE. However now that many traders active on major markets are no longer humans but instead are high-speed automated systems, transmission latencies can matter a great deal. This is just one respect in which BSE is a major simplification of the real-world situation. Real-world exchanges and markets are really *much* more complicated than BSE. Nevertheless, the abstractions embodied within BSE render it a genuinely useful platform for leading-edge research.

BSE is written in *Python v.2.7* as a single-threaded process intended to be run in batch-mode, writing data to files for subsequent analysis, rather than single-stepping with dynamic updating of displays via an interactive graphical user interface (GUI). A GUI-based version is in development, but even that would require the user to switch into batch-mode to generate statistically reliable data from many hundreds or thousands of repeated market simulation experiments, as is discussed further in Section 3.3.

For ease of distribution, and to help people who are new to *Python*, currently all of BSE fits in a single source-code file: BSE.py, the latest major release of which is available for download from the popular GitHub open-source repository (BSE, 2012). BSE.py has been written to be easy to understand; it is certainly not going to win any prizes for efficiency; probably not for elegance either. It's roughly 1000 lines of code.

The output data-files created by BSE.py are all ASCII comma-separated values (i.e., files of type .csv) because that format can easily be imported by all popular spreadsheet programs (such Microsoft *Excel*) and can also be readily imported by more sophisticated statistical analysis systems such as those offered by *Matlab* or *R*.

Examining the BSE code you can see that the Exchange has to keep internal records of exactly which trader submitted which order, so that the book-keeping can be done when two traders enter into a transaction, but that the LOB it "publishes" to the traders deliberately discards a lot of that data, to anonymize the identity of the traders. This is exactly what real-world LOB-based exchanges do. The Exchange class's `publish` method uses values from the exchange's internal data structures to build the market's LOB data as a *Python* dictionary structure containing the time, the bid side of the LOB and the ask side of the LOB, referred to as the Bid LOB and the Ask LOB, respectively. The Bid and Ask LOBs are both also dictionary structures. Each LOB shows: the current best price; the worst possible price (i.e. the lowest-allowable bid-price, or the highest-allowable offer price: these values can be of use to trader algorithms, i.e. for making "stub quotes": an example is given later in this document); the number of orders on the LOB; and then the anonymized LOB itself.

The anonymized LOB is a list structure, with the bids and asks each sorted in price order. Each item in the list is a pair of items: a price, and the number of shares bid or offered at that price. Prices for which there are currently no bids or asks are not shown on the LOB.

Quotes that are issued by a trader have a trader-identification (TID) code, a quote type (bid or ask), price, quantity, and a timestamp. In the current version of BSE, for extra simplicity, the quantity is always 1. In this document we'll show the quote as a list: [TID, type, price, quantity, time], hence if trader T22 bids $1.55 for one share at time *t*=10 seconds after the market opens, we'd write the quote as: `[T22,bid,155,1,0010]`. Figure 2 shows an example in which the LOB is initially empty and on successive lines the LOB is shown after it is updated in response to the quote shown on that line.

Note that in Figure 2 the order issued at *t*=21 comes from trader T11, and hence replaces T11's previous order issued at *t*=2, which is why the bid at $0.22 is deleted from the bid LOB at *t*=21. At this stage, the *bid-ask spread* (i.e., the difference between the best bid and the best ask) is $0.32.

```
                           bid:[]              ask:[]
[T11, bid, 022, 1, 0002]   bid:[(22,1)]        ask:[]
[T02, bid, 027, 1, 0006]   bid:[(22,1),(27,1)] ask:[]
[T08, ask, 077, 1, 0007]   bid:[(22,1),(27,1)] ask:[(77,1)]
[T01, bid, 027, 1, 0010]   bid:[(22,1),(27,2)] ask:[(77,1)]
[T03, ask, 062, 1, 0018]   bid:[(22,1),(27,2)] ask:[(62,1),(77,1)]
[T11, bid, 030, 1, 0021]   bid:[(27,2),(30,1)] ask:[(62,1),(77,1)]
```

Figure 2: Changes in the LOB Data Structure in Response to a Succession of Orders. The top line shows the initially empty LOB, as the bid-side list (center column) and the ask-side list (right-hand column). Successive lines show a sequence of orders arriving (left-hand column) and the resultant state of the LOB after it is updated to represent each order: see text for further discussion.

In financial-market terminology, if a trader wants to sell at the current best bid price, that's referred to as "hitting the bid"; if a trader wants to buy at the current best ask-price, that's referred to as "lifting the ask". Both are instances of "crossing the spread". So, to continue our example, let's say that trader T02 decides to lift the ask.

In BSE, this is signalled by the trader issuing an order that crosses the spread, i.e. issuing a bid priced at more than the current best ask, and the transaction then goes through at whatever the best price was on the LOB as the crossing order was issued. So, continuing the sequence of events in Figure 2, if at *t*=25s Trader 02 lifts the ask by bidding $0.67, the updated LOB would be as illustrated in Figure 3.

`[T02, bid, 067, 1, 0025]  bid:[(27, 1),(30, 1)] ask:[(77, 1)]`

Figure 3: Illustration of changes in the LOB in response to a quote issued at t=25s, one that crosses the spread, thereby lifting the best ask.

Note that, as shown in Figure 2, T02 already had an earlier bid on the LOB, one of the two priced at 27, so when T02's new bid of 67 was received at *t*=25s by BSE, it first deleted the earlier bid (i.e., the new order replaced the old one), and then the exchange system detected that the bid crossed the best ask (which was priced at 62, from the order sent at *t*=18s by trader T03) and so the exchange deleted that ask from the LOB too. Immediately afterwards, the exchange and the traders do some book-keeping: the exchange records the transaction price and time on its "tape" (the time-series record of orders on the exchange); and the two counterparties to the trade, traders T02 and T03, each update their "blotters" (their local record of trades they have entered into): these uses of the words "tape" and "blotter" are common terminology among financial-market traders.

### 3.2 BSE Robot Traders

BSE includes a sample of various simple automated trading algorithms, or "robot traders". These are all automated execution systems: that is, they automate the process of executing an order that has originated elsewhere. In investment banking, the human workers that do this job are known as "sales traders": a human sales trader waits for an order to arrive from a customer, and then works that order in the market: that is, the sales trader just executes the order without giving any advice or comment on the customer's decision to buy or sell. At its simplest, a customer order will state what instrument (e.g., which stocks or shares) the customer wants to deal in, whether she wants to buy or sell, how many, and what price she wants. If the customer is keen to complete the transaction as soon as possible, she should instead specify a *market order*, i.e. just do the deal at whatever is the best price in the market right (i.e., on the LOB) now. But if the customer is happy to wait, then she can specify a *limit order*, a maximum price for a purchase or a minimum price for a sale, and then the sales trader's job is to wait until the conditions are right for the deal to be done. The execution of limit orders is where the sales trader can make some money. Say for example that the customer has sent an order stating that she wants to sell one share of company XYZ for no less than $10.00, at a time when XYZ is trading at $9.90 but where the price has been rising steadily: if the sales trader waits a while and executes the order when XYZ has risen to $10.50, the trader can return the $10.00 to the customer and take the extra $0.50, a margin of 5%, as a fee on the transaction; but if the trader instead executes this order at the precise moment that the price of XYZ hits $10.00, then the customer's order has been satisfied but there is no extra money, no margin on the deal, for the trader to take a share of. Similarly for a customer order to buy at a limit price of than $15.00, if the trader can instead execute the order by buying at $13.00 then there is $2.00 "profit" to keep or share, a margin of 13%.

So, the simple robot traders in BSE can all be thought of as computerized sales-traders: they take customer limit orders, and do their best to execute the order in the market at a price that is better than the limit-price provided by the customer (this is the price below which they should not sell, or above which they should not buy). Customer orders are issued to traders in BSE and then the traders issue their own quotes as bids or asks into the market, trying to get a better price than the limit-price specified by the customer.

BSE is written as object-oriented *Python*. There is a generic Trader class that specifies stub implementations of core methods that any robot needs to implement such as how the robot calculates the price of its next quote that it will send as an order to the LOB (`getorder()`), how to update any learning variables in response to market events (`respond()`), or the book-keeping and record-updating that needs to be done when an order is executed (`bookkeep()`). The definitions of each specific type of robot trader in BSE then inherit these generic methods and extend them as required.

For brevity in the BSE code each type of robot trader has an identifier of up to four characters, similar to a stock-ticker symbol. The robot trader algorithms currently available in BSE are:

- **Giveaway (`GVWY`):** a totally dumb robot that issues a quote-price that is identical to its limit price, thereby maximising its chances of finding another trader to transact with, but guaranteeing to make no profit should a trade result from its quote. The GVWY trader makes no use of any LOB data.
- **Zero-Intelligence Constrained (`ZIC`):** an implementation of Gode & Sunder's ZIC traders, as introduced in their seminal 1993 paper which demonstrated that, when evaluated via a common measure of the efficiency of market mechanisms, markets populated by ZIC traders are just as efficient as comparable markets populated by human traders.
- **Shaver (`SHVR`):** this is a minimally simple trader that, unlike GVWY, does actually use LOB data. If it is working a sell order, the SHVR algorithm simply looks at the best ask on the LOB and undercuts it by quoting one penny less (i.e., 0.01: the smallest unit of currency in BSE) so long as this does not go below the sell order's limit price. Similarly if working a buy order, SHVR quotes a bid-price that is one penny more than the current best bid, so long as that price is not more than the buy-order's limit price.

- **Sniper (SNPR):** in a famous early public contest in automated trading on experimental markets, organized at the Santa Fe Institute, Todd Kaplan submitted a trader-robot spent most of its time doing nothing, "lurking in the background" and then, if the market was about to close or if the bid-ask spread had narrowed to a sufficiently small value, Kaplan's strategy came in to the market to "steal the deal". This strategy, now known as *Kaplan's Sniper*, won the contest: see Rust *et al.* (1992) for further details. In BSE the code for SHVR is extended to include elementary sniping capability: the BSE SNPR robot reads a system-wide global variable that indicates the percentage of time remaining in the market session, lurks for a while, and then rapidly increases the amount it shaves off the best price as time runs out.
- **Zero-Intelligence Plus (ZIP):** the ZIP trading algorithm (Cliff, 1997) was devised to address shortcomings in ZIC traders. A ZIP trader uses simple machine learning and a shallow heuristic decision tree to dynamically alter the margin that it aims to achieve on the order it is currently working. In 2001, Das *et al.* at IBM reported on experiments in which they demonstrated that ZIP, and also IBM's own "MGD" algorithm modified from an algorithm first reported by Gjerstad & Dikhaut (1998), could outperform human traders in controlled laboratory experiments.
- **Adaptive-Aggressive (AA):** for his PhD research, Vytelingum (2006) made significant extensions to the ZIP algorithm, adding an *aggressiveness* variable that determines how quickly the trader alters its margin, and this variable is itself adaptively altered over time in response to events in the market. AA has been shown to dominate prior trading algorithms such as ZIP and MGD (Vytelingum, Cliff, & Jennings, 2008). In later work (De Luca & Cliff, 2011) AA was demonstrated to dominate not only all other algorithmic trading strategies, but also human traders.

The *Python* source-code for each of these robot traders is available on the BSE GitHub repository (BSE, 2012).

### 3.3 Using the LOB with Robot Traders

The core of a market session in BSE is a while loop that repeats once per time-step until a pre-specified time-limit is reached. In each loop-cycle, the following happens:

- There is a check to see if any new customer orders need to be distributed to any of the traders, via a call to the BSE `customer_orders()` method. The call to that method includes a parameter, `order_schedule,` which specifies various aspects of how customer orders are generated, such as whether new orders arrive randomly or regularly in time, what the balance is between sell orders and buy orders, and how the price of each order is generated. Prices can be constant, or generated according to a deterministic function of time, or generated at random from a stochastic function: conditionally heteroscedastic price-generating functions can easily be constructed.
- An individual trader is chosen at random to issue its current response by invoking the trader's `getorder()` method: this will either return the value `None`, signalling that the trader is not issuing a quote at the current time, or it will return a quote, i.e. a fresh order to be added to the LOB; if that is the case then BSE processes the order via a method called `process_order()`.
- The updated LOB is then made available to all traders via a call to BSE's `publish_lob()` method.
- If processing the order resulted in a trade, the traders involved do the necessary book-keeping, updating their blotters, via a call to `bookkeep()`.
- Each trader is given the chance to update its internal values that affect its trading behavior, via a call to the trader's `respond()` method.

Before any of that happens, the market needs to be populated by an invocation of the `populate_market()` method, which is where the number of traders and the type of each trader is determined. Thus the core loop of the BSE simulator resembles the code shown in Figure 4. To make a proper rigorous evaluation, comparing different trading robots across a realistic variety of market conditions, it is necessary to run multiple sequences of statistically independent sessions, and then calculate appropriate summary statistics from, and/or perform appropriate tests of statistical significance on, the results. This is easily done in BSE. Nevertheless, establishing which robot trader performs best across a number of different supply and demand schedules, and with differing numbers and ratios of robot types, can require very large number of trials, of individual market sessions. In a teaching context, this is an advantage: BSE's ability to routinely generate very large data-sets can be used to motivate students to learn "big data" tools and techniques for managing, visualizing, and analyzing large data-sets.

### 3.4. Altering supply and demand schedules in BSE

De Luca *et al.* (2011) and Cartlidge & Cliff (2012) discuss the need to explore trading agents in simulated markets that are more realistic than those used in prior experimental studies of robot traders interacting with one another and/or with human traders. The current version of BSE can be configured to run "traditional" economics experiments switching between static-equilibrium supply/demand schedules and with periodic simultaneous replenishment of orders, but it can also be configured to have continuous "drip-feed" replenishment, along with fine-grained dynamic variations in the supply and demand schedules, and hence also in the market equilibrium.

BSE allows the customer orders to arrive in a continuous random stream, rather than periodically having every single trader being given a new customer order, all at the same instant. Such full periodic replenishment of customer orders is something that was introduced in Vernon Smith's seminal (and subsequently Nobel Prize winning) experiments reported in (Smith, 1962), which very many experimenters have used since, but there are good reasons for wanting to explore simulation experiments where instead the replenishment is continuous: in very many real-world markets, for much of the time, the flow of orders is a continuous random feed of orders into the market. For more discussion of this, see Cliff & Preist (2001), De Luca *et al.* (2011).

```python
# initialise population of traders to empty
traders = {}
# set clock to zero
time = 0

# create a population of traders
populate_market(n_traders, traders)

while time < endtime:

    # how much time left, as a percentage?
    duration = float(endtime-starttime)
    time_left = (endtime - time) / duration

    # distribute any new customer orders to the traders
    customer_orders(time, traders, order_schedule)

    # get an order (or None) from a randomly chosen trader
    # tid is Trader Identifier
    tid = random_tid(traders)
    lob = exchange.publish_lob(time)

    order = traders[tid].getorder(time,time_left,lob)

    if order != None:

        # send order to exchange
        trade = exchange.process_order(time, order)

        lob = exchange.publish_lob(time)

        if trade != None:
            # trade occurred,
            # counterparties update order-lists & records
            traders[trade['party1']].bookkeep(trade, order)
            traders[trade['party2']].bookkeep(trade, order)
            # update exchange records/stats for later viz/analysis
            trade_stats(expid, traders, tdump, time, lob)

    # traders respond to whatever happened
    lob = exchange.publish_lob(time)
    for t in traders:
        traders[t].respond(time, trade, lob)

    time = time + timestep
```

Figure 4: Main loop of a market session in BSE. See text for further explanation.

## 4. BSE IN USE

BSE was created in 2012 to meet a need in our teaching of a unit/module currently known *as Internet Economics and Financial Technology*, available to masters-level students at the University of Bristol. In the six years since then approximately 250 masters students have used BSE in coursework assignments, typically requiring the students to develop and test their own robot trader. In some years the assignment required students to create a sales-trader robot (as described above) and in other years a *proprietary-trader* robot, which starts with a sum of money and then buys and sells on its own account, attempting to make a profit by selling each thing it buys for more than the price at which it was purchased. Student feedback on using BSE was generally highly positive, and several graduates of the module have gone on to permanent employment with major investment banks and hedge funds. For more detailed discussion of the use of BSE in teaching see (Cliff, 2018).

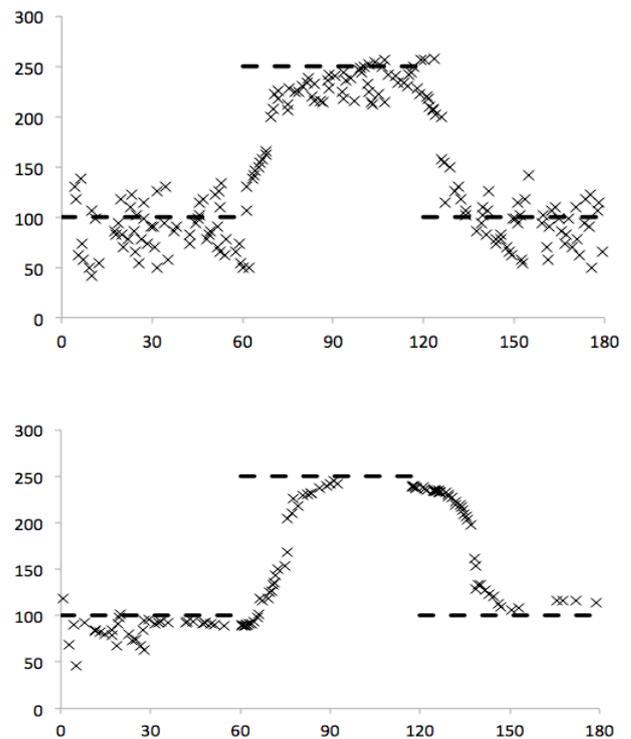

Figure 5: Sample transaction-price time-series from BSE markets with shock-changes in supply and demand. altering the market's equilibrium price. Horizontal axis is time in seconds; vertical axis is price. Data-points are individual transactions. The equilibrium price $P_0$ is indicated by the horizontal dashed line: in the first minute $P_0=\$1.00$, then at $t=60$ there is a shock change to $P_0=\$2.50$, and then after a further minute the equilibrium price undergoes another shock change at $t=120$ back to $P_0=\$1.00$. In each time-series, the traders' response to the step-change in the equilibrium price as it shifts up at $t=60$ and back down at $t=120$ is clear. In each market there are 40 buyers and 40 sellers: upper graph shows results from a market of GVWY traders; lower graph shows results from a market of ZIP traders.

Latterly BSE has been used as the basis for research work, most recently described in (le Calvez & Cliff 2018), exploring the use of deep learning neural networks (DLNNs: see e.g. Krizhevsky *et al.*, 2012) to replicate the behavior of adaptive traders in a CDA markets. DLNNs are a contemporary machine learning technique that have been successfully demonstrated to perform surprisingly well in a wide range of highly challenging application areas, but which typically only perform well when trained on data-sets that are extremely large by traditional standards. BSE can readily generate data in the volumes required for DLNN training, as was first demonstrated by Tibrewal (2017) who showed successful results from training a DLNN network to replicate the trading activity of a specific ZIP robot-trader in BSE, thereby establishing a proof-of-concept that DLNNs could learn to trade in LOB-based CDA markets; le Calvez (2018) recently replicated Tibrewal's work and extended it by demonstrating a DLNN-trained robot trader operating successfully, trading live in BSE, and at times outperforming the trader from which the training data was generated: see le Calvez & Cliff (2018) for further details.

To illustrate this, Figure 6 shows the top-level *Python* code required to launch a sequence of 22,750 successive independent experiments in BSE, each lasting 5 minutes of simulated time, with varying proportions of ZIP, ZIC, GVWY, and SHVR robots, keeping the total number of traders in the market fixed at 32 (16 working buy orders and 16 working sell orders), recording 50 separate experiments for every possible permutation ranging from homogenous markets where all 32 traders are the same type, through to the case where there are 8 traders running each of the four robot strategies. To specify this takes less than 40 lines of code: students interested in studying market dynamics, or evaluating the performance of their own self-0designed robot trading algorithm can operate at the level of code shown in Figure 6: they do not need to understand implementation details of the BSE LOB or the array of robot traders already coded in; all those details are abstracted away and the students need only understand how to interact with the BSE LOB's interface, and the interfaces of the generic `Trader` object class.

## 5. DISCUSSION AND FURTHER WORK

A quick trawl of the Web reveals a variety of stock-market simulators, many of them aimed at day-traders wanting to evaluate an automated strategy. Such simulators almost always are designed to work with historical data of stock prices: by back-testing on historical data, it is possible to estimate how much money an automated trading strategy would have made (or lost) if it had been running live.

Clearly, BSE is quite different to such trading simulators, but there are good reasons for that. Very few trading simulators work with Level 2 data (i.e. with the full LOB, changing order-by-order), and the cost of obtaining such data is often very high. More fundamentally, trading simulators based on historical market data typically cannot model *market impact*, where buying or selling large amounts of an instrument shifts the demand and/or supply curves in such a way that the equilibrium price of the instrument then alters. In that sense, conventional trading simulators require the trader (human or automated) to act solely as a *price-taker*, trading at time $t$ at whatever price is on the screen, whatever price the historical data says the instrument was trading at, at time $t$. Whether they sell 1 share or 10 million shares at time $t$, the price immediately after the sale will still be whatever the historical data says it was at time $t+1$. Yet a sale of 10 million shares would in reality almost definitely move the price down, in a way that selling a single share simply wouldn't. Unlike trading robots dealing with historical data, trading robots in BSE can be *price-makers*: their activity can shift the supply and demand, and they can generate, and have to deal with, market impact. In that sense, the scenario in BSE is more like a modern-day "dark pool", a private online exchange where a relatively small number of traders meet to conduct big transactions: participants in dark pools are typically traders working for major banks or fund-management companies, dealing large blocks of tradable instruments.

Nevertheless, in comparison to real markets, the lack of any latency in the system is the probably the biggest issue. It would be relatively easy to introduce simulated latency at the exchange (so the LOB data that is received by the traders at time $t$ actually reflects the state of the LOB a little earlier, at time $t-\Delta t$) and we could also introduce "communications latency" so that when a trader issues an order at time $t$ it does not arrive at the exchange until a little later at some time $t+\Delta t$, but in a single-threaded simulation it would require quite a lot more work to accurately model processing latency in each trader. That is, in the current version of BSE, each trader gets as long as it wants to process its `respond()` method, whereas in a reality a lot of effort goes into making the response-time of automated trading systems as low as possible while still being capable of generating profitable behaviors. To better model real market systems, we would need to switch to a multi-threaded implementation, and/or to configuring BSE as a distributed client-server architecture over multiple virtual or physical machines. That remains one next step for further work.

## ACKNOWLEDGMENTS
The Python code for Vytelingum's (2006) AA robot-trader strategy was added to the BSE *GitHub* repository by Dr Ash Booth, who at the time was at the University of Southampton and is now Head of Artificial Intelligence at HSBC Bank; I am very grateful to Ash for his contribution to BSE. Thanks also to all the other people who have contributed to BSE on GitHub via their pull requests, edits, and forks.

```python
start_time = 0
start_time = 240

n_trader_types = 4
equal_ratio_n = 4
n_trials_per_ratio = 50
n_traders = n_trader_types * equal_ratio_n

fname = 'balances_%03d.csv' % equal_ratio_n

tdump=open(fname,'w')

min_n = 1

trialnumber = 1
trdr_1_n = min_n
while trdr_1_n <= n_traders:
    trdr_2_n = min_n
    while trdr_2_n <= n_traders - trdr_1_n:
        trdr_3_n = min_n
        while trdr_3_n <= n_traders - (trdr_1_n + trdr_2_n):
            trdr_4_n = n_traders - (trdr_1_n + trdr_2_n + trdr_3_n)
            if trdr_4_n >= min_n:
                buyers_spec = [('GVWY', trdr_1_n),
                               ('SHVR', trdr_2_n),
                               ('ZIC',  trdr_3_n),
                               ('ZIP',  trdr_4_n)]
                sellers_spec = buyers_spec
                traders_spec = {'sellers':sellers_spec,
                                'buyers':buyers_spec}
                print buyers_spec
                trial = 1
                while trial <= n_trials_per_ratio:
                    trial_id = 'trial%07d' % trialnumber
                    market_session(trial_id,
                                   start_time, end_time,
                                   traders_spec, order_sched,
                                   tdump, False)
                    tdump.flush()
                    trial = trial + 1
                    trialnumber = trialnumber + 1
            trdr_3_n += 1
        trdr_2_n += 1
    trdr_1_n += 1
tdump.close()

print trialnumber-1
```

Figure 6: Top-level *Python* code for running an experiment with 32 traders (16 buyers and 16 sellers) of types GVWY, SHVR, ZIC, and ZIP; with the ratio of the four types of traders being systematically varied across all possible nonzero values, and performing 50 independent trials for each specific ratio. The main BSE loop illustrated in Figure 5 is her wrapped into a single method invoked as `market_session()`. The nested loops here cause a total of 22,750 trials to be performed, and on a single CPU would take several hours of continuous computation. As each trial is independent, this task is *embarrassingly parallelizable*: in principle, 22,750 separate machines (e.g., that number of virtual machines/instances rented on a pay-by-the hour basis from a cloud service provider) could perform the necessary computation in parallel, taking only a few seconds.

**REFERENCES**

BSE, 2012. GitHub open-source code repository at http://github.com/davecliff/BristolStockExchange/

Cartlidge, J. & Cliff, D., 2012. Exploring the 'robot phase transition' in experimental human-algorithmic markets. UK Government Office for Science, *Foresight Project: Future of Computer Trading in the Financial Markets, Driver Review DR 25*. https://bit.ly/2llHjbh.

Cliff, D., 1997. Minimal-Intelligence Agents for Bargaining Behaviors in Market-Based Environments. HP Labs Tech Report HPL-97-91. www.hpl.hp.com/techreports/97/HPL-97-91.pdf

Cliff, D. & Preist C., 2001. Days without end: on the stability of experimental single-period CDA markets. HP Labs Tech. Report HPL-2001-325. www.hpl.hp.com/techreports/2001/HPL-2001-325.pdf

Cliff, D., 2018. A Free Open-Source Limit-Order-Book Simulator for Teaching and Research. Submitted to *Computational Intelligence in Financial Engineering (CIFEr)* track at IEEE Symposium Series on Computational Intelligence, Bengaluru, India, November 2018.

Das, R., Hanson, J., Kephart, J., & Tesauro, G., 2001. Agent-Human Interactions in the Continuous Double Auction. *Proceedings International Joint Conference on Artificial Intelligence* (IJCAI'01).

le Calvez, A., 2018. *Learning to be a Financial Trader: An Exploration of Neural Networks in a Continuous Double Auction.* Masters Thesis, Department of Computer Science, University of Bristol.

le Calvez, A. & Cliff, D., 2018. Deep Learning can Replicate Adaptive Traders in a Limit-Order-Book Financial Market. Submitted to *Computational Intelligence in Financial Engineering (CIFEr)* track at IEEE Symposium Series on Computational Intelligence, Bengaluru, India, November 2018.

De Luca, M., & Cliff, D., 2011. Human-Agent Auction Interactions Adaptive-Aggressive Agents Dominate. *Proceedings International Joint Conference on Artificial Intelligence* (IJCAI-2011).

De Luca, M., Cartlidge, J., Szostek, C., & Cliff, D., 2012. Studies of interactions between human traders and algorithmic trading systems. UK Government Office for Science, *Foresight Project: Future of Computer Trading in Financial Markets, Driver Review DR13*. https://bit.ly/2llv52c.

De Luca, M., 2016. *Adaptive Algorithmic Trading Systems.* PhD thesis, Department of Computer Science, University of Bristol.

Gjerstad, S. & Dickhaut, J., 1998. Price Formation in Double Auctions. *Games & Economic Behavior*, 22(1):1-29.

Gode, D. & Sunder, S., 1993. Allocative efficiency of markets with zero-intelligence traders: Market as a partial substitute for individual rationality. *Journal of Political Economy*, 101(1):119-137.

Krizhevsky, A., Sutskever, I, & Hinton, G., 2012. ImageNet classification with deep convolutional neural networks. In *Proc. 25th International Conf. on Neural Information Processing Systems, Vol.1 (NIPS'12)*, F. Pereira, *et al.* (eds), pp.1097-1105.

Rust, J., Miller, J., & Palmer, R., 1992. Behaviour of trading automata in a computerized double auction market. in D. Friedman & J. Rust (eds) *The Double Auction Market: Institutions, Theories, & Evidence*. Addison Wesley, pp.155-198.

Smith, V., 1962. An experimental study of competitive market behavior. *J. Polit. Economy*, 70(2):111-137.

Stotter, S., Cartlidge, J., & Cliff, D., 2013. Exploring assignment-adaptive (ASAD) trading agents in financial market experiments, in *Proceedings of the 5th International Conference on Agents and Artificial Intelligence,* (ICAART). J. Filipe & A. Fred, (eds). Barcelona: SciTePress, Vol.1.pp.77-88.

Stotter, S., Cartlidge, J., & Cliff, D., 2014. Behavioural investigations of financial trading agents using Exchange Portal (ExPo). In: N. Nguyen, *et al.* (eds) *Transactions on Computational Collective Intelligence XVII*. Springer, pp. 22-45

Tibrewal, K., 2017. *Can Neural Networks be Traders? Explorations of Machine Learning Using the Bristol Stock Exchange*. MEng Thesis, Department of Computer Science, University of Bristol.

Vytelingum, P., 2006, *The Structure and Behaviour of the Continuous Double Auction*. PhD Thesis, School of Electronics and Computer Science, University of Southampton.

Vytelingum, P., Cliff, D., & Jennings, N. 2008. "Strategic Bidding in Continuous Double Auctions". *Artificial Intelligence*, 172(**14**):1700-1729.

**AUTHOR'S BIOGRAPHY**

**Dave Cliff** is a full Professor of Computer Science at the University of Bristol, UK, a post he has held for the past 11 years. Previously he has held professorial posts at the University of Southampton, UK, and at the MIT Artificial Intelligence Lab in Cambridge, USA. He also spent seven years working in industrial technology R&D at Hewlett-Packard Laboratories and as a Director & Trader in the Foreign Exchange Complex Risk Group at Deutsche Bank in the City of London. He has a BSc in Computer Science (Leeds, 1987) and MA and PhD degrees in Cognitive Science (Sussex, 1988 & 1992). For the past 22 years Cliff's research has focused on applications of artificial intelligence and machine learning in financial markets, and on issues in systemic risk and stability in networked financial systems. Cliff has recently served as a lead expert advisor to the UK Government Office for Science in its two-year "Foresight" investigation *The Future of Computer Trading in Financial Markets*, for its *Blackett Review* of the UK FinTech Industry, and is a founder member the Academic Advisory Council on Data Analytics for the UK Financial Conduct Authority, the main regulator of financial-market activity in the UK. He is a Fellow of the British Computer Society and a Fellow of the Royal Society of Arts.